\documentclass[12pt,nofootinbib]{revtex4}
\usepackage{graphicx}
\usepackage{float}
\usepackage{color}
\usepackage{ulem}
\usepackage{hyperref}

\newcommand{\be}{\begin{equation}}
\newcommand{\bea}{\begin{eqnarray}}
\newcommand{\beq}[1]{\begin{equation}\label{#1}}
\newcommand{\beqa}[1]{\begin{eqnarray}\label{#1}}
\newcommand{\eq}[1]{Eq.~(\ref{#1})}
\newcommand{\eqs}[2]{Eqs.~(\ref{#1}) and~(\ref{#2})}

\newcommand{\ee}{\end{equation}}
\newcommand{\eea}{\end{eqnarray}}
\newcommand{\eeq}{\end{equation}}
\newcommand{\eeqa}{\end{eqnarray}}

\def\gev{\,{\rm GeV}}

\newcommand{\diff}{\mathrm{d}}
\newcommand{\Ord}[1]{\mathcal{O}\left(#1\right)}
\newcommand{\msbar}{\overline{\text{MS}}}
\newcommand{\Obs}[2]{\widehat{O}_{#1}^{\text{#2}}}

\newcommand{\Olow}{\{\widehat{O}_i^{\text{low}}\}}
\newcommand{\Ohigh}{\{\widehat{O}_i^{\text{high}}\}}
\newcommand{\Oin}[1]{\{\widehat{O}_{#1}^{\text{in}}\}}
\newcommand{\mQ}{\{m_Q(m_Q)\}}
\newcommand{\pother}{\{p_k^{\text{other}}\}}
\newcommand{\inp}{\{\mathcal{I}_k\}}
\newcommand{\GC}{\left\langle\frac{\alpha_s}{\pi}G^2\right\rangle}

\newcommand{\alsmz}{\alpha_s(m_Z)}
\newcommand{\almz}{\alpha(m_Z)}
\def\mom{{\cal M}}
\newcommand{\mua}{\mu_\alpha}
\newcommand{\mum}{\mu_m}
\newcommand{\mumin}{\mu_{\rm min}}
\newcommand{\mumax}{\mu_{\rm max}}
\newcommand{\Gamc}{\Gamma_{H\to c\bar c}}
\newcommand{\Gamb}{\Gamma_{H\to b\bar b}}

\begin{document}

\begin{flushright}
\begin{small}
FERMILAB-Pub-15-001-T \\ WSU-HEP-1501 \\ MCTP-15-01
\end{small}
\end{flushright}

\title{The role of low-energy observables in precision Higgs analysis}
\author{Alexey~A.~Petrov$^{a,b,c}$, Stefan~Pokorski$^d$, James~D.~Wells$^b$ and Zhengkang~Zhang$^b$}
\address{$^{(a)}$Department of Physics and Astronomy, Wayne State University, Detroit, MI 48201, USA\\
$^{(b)}$Michigan Center for Theoretical Physics, Department of Physics, University of Michigan, Ann Arbor, MI 48109, USA\\
$^{(c)}$Theoretical Physics Department, Fermilab, P.O. Box 500, Batavia, IL 60510, USA \\
$^{(d)}$Institute of Theoretical Physics, University of Warsaw, Pasteura 5, 02-093 Warsaw, Poland}
\date{\today}

\begin{abstract}  
A conventional approach to precision calculations of Higgs boson observables uses quark masses $m_c$ and $m_b$ as inputs. However, quark masses are single numbers that hide a variety of low-energy data from which they are extracted, and also hide the various sources of theoretical uncertainties and correlations with additional input parameters such as $\alpha_s$. Higher-precision calculations, which are needed to give meaning to future measurements, require more direct engagement with the low-energy data in a global analysis. We present an initial calculation in this direction, which illustrates the procedure and reveals some of the theory uncertainties that challenge subpercent determinations of Higgs boson partial widths.
\end{abstract}  

\maketitle


\vfill\eject




\section{Introduction}

The discovery of the Higgs boson~\cite{Aad:2012tfa,Chatrchyan:2012ufa} marks the beginning of a new era for precision studies. Not only is unprecedented precision achieved in Standard Model (SM) calculations~\cite{Eberhardt:2012gv,Ciuchini:2013pca,Wells:2014pga,Baak:2014ora} with the knowledge of the Higgs boson mass~\cite{Chatrchyan:2012jja,Aad:2013wqa}, but experimental data on a large number of Higgs observables~\cite{ATLAScouplings:2014} allows us for the first time to scrutinize the Higgs sector of the SM~\cite{Almeida:2013jfa} and beyond~\cite{Englert:2014uua,Djouadi:2013lra,Contino:2013kra}. Any discrepancy between precision data and SM predictions would be an indication of new physics.

Though not explicitly stated in the context of precision Higgs analysis, an important role in this program is played by low-energy observables, such as moments of $e^+e^-$ annihilation cross section and moments of semileptonic $B$ decay distributions. In fact, our knowledge of the charm and bottom quark masses $m_Q$ ($Q=c,b$), which are important inputs of precision Higgs calculations, largely comes from analyzing these low-energy data. This can be seen from the fact that the Particle Data Group (PDG)~\cite{Agashe:2014kda} average of the scale-invariant masses in the $\msbar$ scheme [i.e.~solutions to $m_Q(\mu)=\mu$],
\bea
m_c(m_c) &=& 1.275(25)\text{ GeV},\label{mcpdg}\\
m_b(m_b) &=& 4.18(3)\text{ GeV},\label{mbpdg}
\eea
is dominated by $m_Q$ extractions from low-energy data. These $\msbar$ masses, as well as pole masses, have been used in the literature to estimate the theoretical precision achievable in precision Higgs calculations~\cite{Heinemeyer:2013tqa,Almeida:2013jfa}.

However, looking into the future, such indirect engagement of low-energy observables in precision Higgs analysis might be ultimately unsatisfactory. A large amount of low-energy data has been highly processed to yield just two numbers, as in \eqs{mcpdg}{mbpdg}. It is not even clear whether these numbers accurately reflect our knowledge of $m_Q$, because the averaging involves $m_Q$ extractions some of which are apparently correlated due to similar data and/or methods used. The error bars assigned to them contain experimental uncertainties from many different measurements, as well as theoretical uncertainties from calculating many different quantities. In addition, a self-described inflation of uncertainties by the PDG~\cite{pdgMQ} is introduced to account for underestimated systematic errors in some $m_Q$ extractions~\cite{Dehnadi:2011gc}. Finally, \eqs{mcpdg}{mbpdg} do not retain possible correlations between $\alsmz$ and the extracted $m_Q$. They are thus treated as independent inputs in precision Higgs analysis, which is strictly speaking not correct.

As we strive for the highest-precision calculation possible in order to match percent (or even perhaps parts-per-mil) level of experimental precision achievable in the foreseeable future\footnote{Though precision measurements of Higgs observables, especially the partial widths into $c\bar c$ and $b\bar b$ discussed in this paper, are difficult at the LHC, such high precision is generally believed to be achievable at the International Linear Collider, the Future Circular Collider, and the Circular Electron Positron Collider. For recent analyses, see e.g.~\cite{Asner:2013psa,Peskin:2013xra,Fan:2014vta,Ruan:2014xxa}. We also note that for the $b\bar b$ channel, the importance of a higher theory precision is further emphasized by its relevance to the calculation of the total widths and all branching ratios of the Higgs boson.}, the rich information hidden in \eqs{mcpdg}{mbpdg} should be revealed, and the role of individual low-energy observables emphasized. Conceivably, a global $\chi^2$ fit would become more powerful in testing the SM when low-energy observables sensitive to $m_Q$ as well as Higgs observables are incorporated. The scale-invariant masses $m_Q(m_Q)$ would be then only inputs of the {\it calculation}. They are not considered as {\it observables} with experimental values and uncertainties, but are parameters to be tuned to minimize the $\chi^2$ function, where only true {\it observables} are included.

In this paper we propose the idea of directly working with low-energy observables in precision Higgs analysis. In addition to the global fit perspective mentioned above, low-energy observables can also play a role in identifying individual sources of theoretical uncertainties in precision Higgs calculations. This is conveniently done by eliminating $m_Q(m_Q)$ from our input in favor of two low-energy observables, and recasting Higgs observables in terms of these and other input observables. For this procedure to be meaningful, the two observables chosen should be representative of the large amount of low-energy data contributing to \eqs{mcpdg}{mbpdg}, in the sense that $m_Q$ extracted from them alone should be precise enough. In the language of a global $\chi^2$ fit, the ideal choices would be two observables that dominate the low-energy observables contribution to $\chi^2$. In this regard, a reasonable, though by no means exclusive, option would be to use the moments 
$\mom_1^c$ and $\mom_2^b$ of $e^+e^-\to Q\bar Q$ inclusive cross section, defined by
\beq{momdef}
\mom_n^Q \equiv \int\frac{\diff s}{s^{n+1}}R_Q(s),\quad\text{where}\,\, R_Q\equiv\frac{\sigma(e^+e^-\to Q\bar QX)}{\sigma(e^+e^-\to \mu^+\mu^-)},
\eeq
with the precise definition of $R_Q$ from experimental data discussed in~\cite{Kuhn:2007vp}. $m_c(m_c)$ and $m_b(m_b)$ reported in the literature from analyzing these moments typically have $\Ord{10~\text{MeV}}$ uncertainties quoted~\cite{Kuhn:2001dm,Kuhn:2007vp,Chetyrkin:2009fv,Dehnadi:2011gc}. For the Higgs observables we will focus on the partial widths $\Gamc$ and $\Gamb$, and assess the level of precision we can achieve in SM predictions for them. We will see that with direct contact made between these partial widths and the low-energy moments, the vague notion of ``uncertainties from $m_Q$'' is decomposed into concrete sources of uncertainties. In particular, parametric uncertainties from input observables $\mom_1^c$, $\mom_2^b$ and $\alsmz$ \footnote{It should be noted that we will treat $\alpha_s(m_Z)$ as both a calculational input and an observable with a central value and uncertainty. In principle one could treat $\alpha_s(m_Z)$ as merely a calculational parameter and let the observables that are highly sensitive to the $\alpha_s(m_Z)$ value be part of the global fit, analogous to what we have done with $m_Q(m_Q)$. However, $\alpha_s(m_Z)$ is one step further removed from direct determination of $H\to b\bar b$, $c\bar c$ partial widths compared to $m_Q(m_Q)$, and so treating $\alpha_s(m_Z)$ as both an input parameter and (highly processed) observable is numerically justified.}, and perturbative uncertainties due to missing higher-order corrections to the moments can be exposed separately. We note that while the parametric uncertainties are currently expected to be at the percent level, and are in principle reducible with future data and more careful experimental extraction of the moments, the perturbative uncertainties may represent a bigger challenge due to lack of knowledge of the appropriate renormalization scales in the low-energy regime. It is therefore worthwhile to further investigate theoretical as well as experimental aspects of the low-energy observables for the precision Higgs program to succeed.

\section{Incorporating low-energy observables into a global precision analysis}

The strongest tests of the SM rely on comparing its predictions across all accessible energy scales. 
By disentangling the information contained in the charm and bottom quark masses in the context of precision Higgs analysis, we expose an interesting interplay between Higgs observables and low-energy observables. The sensitivity to $m_Q$ that they share in common suggests the inclusion of both in the precision program.

An incomplete list of candidates for low-energy observables can be inferred from the $m_Q$ extraction literature, and includes low~\cite{Kuhn:2001dm,Kuhn:2007vp,Chetyrkin:2009fv,Dehnadi:2011gc} and high~\cite{Signer:2008da,Hoang:2012us,Penin:2014zaa,Beneke:2014pta} moments of $R_Q$ mentioned above, and their variants~\cite{Bodenstein:2011ma,Bodenstein:2011fv}, moments of lepton energy and hadron mass distributions of semileptonic $B$ decay~\cite{Gambino:2013rza,Buchmuller:2005zv,Bauer:2004ve},~etc. We denote them collectively as $\Olow$, with $i$ running from 1 to the number of low-energy observables we wish to incorporate into the analysis. All these candidates should be carefully examined, and correlations among them should be understood, so that the best choices can be made for $\Olow$. 

In the high-energy regime, the observables include, for example, various partial widths, branching ratios, and production cross sections of the Higgs boson. Let us call them $\Ohigh$. If not restricted to precision Higgs analysis, one may even include in $\Ohigh$ the electroweak observables, such as the effective weak mixing angle, $Z$ boson partial widths, and forward-backward asymmetries in $e^+e^-$ annihilation at the $Z$ pole. This will make the global analysis even more powerful, because the Higgs observables are sensitive to the same set of input observables as the electroweak observables:
\be
\Oin{k} \equiv \{ m_Z,\; G_F,\; \almz,\; m_t,\; \alsmz,\; m_H \}.
\ee
Parenthetically we remark that the common practice of treating the top quark mass $m_t$ as an input observable is justified for present purposes. A more careful treatment of $m_t$, like what we do here with $m_c$ and $m_b$, may be needed in the future when precision measurements on the $t\bar t$ threshold are carried out at an $e^+e^-$ collider.

Additional calculational inputs, which are not necessarily of the observable type, include the charm and bottom quark masses $\mQ\equiv\{m_c(m_c),m_b(m_b)\}$. There may be other input parameters, which we denote collectively by $\pother$. Examples are the $\tau$ lepton mass, flavor angles, and nonperturbative parameters (e.g.~gluon condensate) involved in some low-energy observables.

Assuming the potentially complicated correlations among all the high- and low-energy observables will be understood in time, we may ultimately subject all the observables to a global fit, by minimizing the $\chi^2$ function with respect to the inputs:
\bea
\text{Calculation inputs:}\quad & & \inp \equiv \Oin{k}\cup\mQ\cup\pother,\\
\text{Fit observables:}\quad  & &\{\Obs{i}{}\}\equiv \Oin{i}\cup\Ohigh\cup\Olow,\\
\text{To minimize:}\quad & & \chi^2 = \sum_{ij} \biggl[\Obs{i}{th}(\inp) - \Obs{i}{expt}\biggr] V^{-1}_{ij} \biggl[\Obs{j}{th}(\inp) - \Obs{j}{expt}\biggr].
\eea
Here ``th'' and ``expt'' denote theoretical and experimental values, respectively, and $V$ is the covariance matrix containing uncertainties and correlations among observables. The calculational inputs could just as well be chosen to be a minimal set of Lagrangian parameters; however, it is most convenient for our purposes to choose a combination of observables and Lagrangian parameters as the minimal set of calculational inputs.

Compared with the conventional approach where low-energy data contribute indirectly via the averaged $\mQ$, our proposal of directly working with low-energy observables allows appropriate treatment of all the correlations and uncertainties. In particular, there is no averaging over correlated $m_Q$ extractions, and the calculational inputs $\mQ$ and $\alsmz$ are no longer correlated. Challenging as it is, such a global analysis is worth further investigation. As a long-term goal for the precision program, it will test our understanding of elementary particle physics at an unprecedented level.

As a final remark in this section, the techniques described above are to be employed in a rigorous test of the SM. The resulting statistical test from the $\chi^2$ analysis is for determining the likelihood of the compatibility of the data with the SM hypothesis. It is straightforward to apply these techniques to a slightly different model, which we call the $\kappa$SM, defined to be exactly the SM theory except that each coupling of the Higgs boson to SM states has a free parameter $\kappa_i$ in front that is varied to fit the data (see e.g.~\cite{Englert:2014uua,Falkowski:2013dza,CMS:2014kappa}). In that case, the $\chi^2$ analysis must include these $\kappa_i$ as {\it extra} input variables and the resulting fit tests the compatibility of the $\kappa$SM theory with the data and, if compatible, gives confidence intervals for the $\kappa_i$ values. Just as with the SM, at the next level of precision analysis of the $\kappa$SM it is important to address the role of low-energy observables that we study in this paper.

\section{Recasting Higgs observables in terms of low-energy observables}

In order to investigate sources of theoretical uncertainties in calculating the Higgs observables, it is helpful to recast them in terms of a set of input observables without invoking a global fit. In the simplest case, suppose all the observables under consideration are insensitive to $\pother$. We choose two low-energy observables $\Obs{1}{low}$, $\Obs{2}{low}$. By inverting the functions
\beq{Olow12}
\Obs{1}{low} = \Obs{1}{low} \Bigl[\Oin{k},\mQ\Bigr],\quad \Obs{2}{low} = \Obs{2}{low} \Bigl[\Oin{k},\mQ\Bigr],
\eeq
we express the quark masses in terms of $\Obs{1}{low}$, $\Obs{2}{low}$:
\beq{mQext}
m_c(m_c) = m_c(m_c) \Bigl[\Oin{k},\Obs{1}{low},\Obs{2}{low}\Bigr],\quad m_b(m_b) = m_b(m_b) \Bigl[\Oin{k},\Obs{1}{low},\Obs{2}{low}\Bigr].
\eeq
$\mQ$ can then be eliminated from the calculation of the Higgs observables:
\beq{Ohighi}
\Obs{i}{high} = \Obs{i}{high} \Bigl[\Oin{k},\mQ\Bigr] = \Obs{i}{high} \Bigl[\Oin{k},\Obs{1}{low},\Obs{2}{low}\Bigr],
\eeq
and we have achieved the goal of recasting Higgs observables in terms of low-energy input observables $\Obs{1}{low}$, $\Obs{2}{low}$. From \eq{Ohighi} it is clear that the precision in the SM prediction for the Higgs observables will benefit from improved knowledge of $m_Q$, which ultimately comes from better measurements of the low-energy observables.

Our choices for the low-energy input observables,
\be
\Obs{1}{low},\Obs{2}{low} = \mom_1^c, \mom_2^b,
\ee
require only a slight generalization of the simple formalism above. We will take into account an additional input, the gluon condensate, as $\pother$ in the case of $\mom_1^c$, but its contribution allows for a simplified treatment. In fact, the simplicity of the analysis is our main motivation for choosing these moments as inputs rather than other low-energy observables which lead to similar level of precision in the extracted $m_Q$. For example, if we were to use semileptonic $B$ meson decay observables (see e.g.~\cite{Gambino:2013rza,Buchmuller:2005zv,Bauer:2004ve}), more input parameters in $\pother$ will show up, including flavor angles and four nonperturbative parameters. Also, the low moments ($\mom_n^Q$ with $n\le4$) chosen here are computationally more straightforward than the high moments ($n\ge10$; see e.g.~\cite{Signer:2008da,Hoang:2012us,Penin:2014zaa,Beneke:2014pta}). The former can be calculated conveniently in the relativistic theory, while a nonrelativistic effective theory treatment is needed for the latter. In addition, since the calculation involves $\msbar$ quark masses, there is no need for introducing other mass schemes. Potentially large uncertainties associated with mass scheme conversion (e.g.~from pole or kinetic masses to $\msbar$ masses), which is needed for some other methods, can thus be avoided. We also note that the approach of extracting $m_Q$ from the low moments was recently recast by the lattice QCD community~\cite{McNeile:2010ji,Colquhoun:2014ica,Chakraborty:2014aca}, and future development in this direction may shed light on the precision Higgs program~\cite{Lepage:2014fla}.

To calculate $\mom_n^Q$, one applies quark-hadron duality~\cite{Novikov:1977dq} to relate the moments~$\mom_n^Q$ to vector current correlators,
\be
\mom_n^Q = \frac{12\pi^2}{n!}\left(\frac{\diff}{\diff q^2}\right)^n\Pi_Q(q^2)\biggr|_{q^2=0},~~{\rm where}
\ee
\be
(q^2g_{\mu\nu}-q_\mu q_\nu)\Pi_Q(q^2) = -i\int\diff^4x\,e^{iq\cdot x}\langle0|Tj_\mu(x)j_\nu^\dagger(0)|0\rangle,
\ee
with $j_\mu$ being the electromagnetic current of $Q$. $\Pi_Q$ can be calculated as an operator product expansion:
\beq{sr1}
\mom_n^Q = \frac{\bigl(Q_Q/(2/3)\bigr)^2}{\bigl(2m_Q(\mu)\bigr)^{2n}} \sum_{i,j} \bar C_{n,i}^{(j)}(n_f) \biggl(\frac{\alpha_s(\mu)}{\pi}\biggr)^i \ln^j\frac{m_Q(\mu)^2}{\mu^2} + \mom_n^{Q,\text{np}},
\eeq
where $Q_Q$ is the electric charge of quark $Q$. As one can see, the values of these moments depend on the quark masses, a fact that QCD sum rules practitioners use to extract quark masses (for reviews see~\cite{Shifman:1998rb,Colangelo:2000dp}). The two terms in Eq.~(\ref{sr1}) come from perturbation theory and nonperturbative condensates, respectively. The perturbative part is known up to $\Ord{\alpha_s^3}$~\cite{Maier:2009fz}, while the gluon condensate contribution, which dominates $\mom_n^{Q,\text{np}}$, has been calculated to next-to-leading order~\cite{Broadhurst:1994qj}. Note that the coefficients $\bar C_{n,i}^{(j)}$ are functions of $n_f$, the number of active quark flavors. The common choices are $n_f=4$ for $Q=c$ and $n_f=5$ for $Q=b$. These are also the numbers of active quark flavors assumed for $\alpha_s(\mu)$ and $m_Q(\mu)$ in \eq{sr1}. $\alsmz$ is defined for $n_f=5$, and should be matched to the 4-flavor effective coupling at the bottom quark threshold before being used in \eq{sr1} for $\mom_n^c$. In our calculations the matching is done assuming 4.2~GeV for both the threshold scale and $m_b(m_b)$, but all the results are found to be insensitive to the details of threshold matching.

$m_Q(\mu)$ are usually extracted by comparing the theoretical calculation with experimental data for $\mom_n^Q$ (see~\cite{Dehnadi:2011gc,Kuhn:2007vp} for technical details). Normally the lowest moment $\mom_1^c$ is taken for the charm quark so as to suppress the nonperturbative contribution to the subpercent level~\cite{Kuhn:2007vp,Chetyrkin:2010ic,Dehnadi:2011gc}. For the bottom quark the gluon condensate can be safely neglected at the present level of precision~\cite{Kuhn:2007vp}, and the second moment $\mom_2^b$ is preferred due to large experimental uncertainty in $\mom_1^b$. We also neglect $\Ord{m_c^2/m_b^2}$ terms in $\mom_2^b$, not explicitly written out in \eq{sr1}, which constitute a tiny contribution~\cite{Kuhn:2007vp}.

It is pointed out in~\cite{Dehnadi:2011gc} that the scales at which $m_Q$ and $\alpha_s$ are renormalized should be considered independently to avoid bias in the uncertainty estimate. \eq{sr1} then should be generalized to
\beq{sr2}
\mom_n^Q = \frac{\bigl(Q_Q/(2/3)\bigr)^2}{\bigl(2m_Q(\mum)\bigr)^{2n}} \sum_{i,a,b} C_{n,i}^{(a,b)}(n_f) \biggl(\frac{\alpha_s(\mua)}{\pi}\biggr)^i \ln^a\frac{m_Q(\mum)^2}{\mum^2} \ln^b\frac{m_Q(\mum)^2}{\mua^2} + \mom_n^{Q,\text{np}}.
\eeq
The coefficients in this equation $C_{n,i}^{(a,b)}$ can be readily derived from $\bar C_{n,i}^{(j)}$ via renormalization group (RG) equations, and numerical results for $n_f=4$ can be found in~\cite{Dehnadi:2011gc}. Due to unknown $\Ord{\alpha_s^4}$ terms, the calculated $\mom_n^Q$ exhibit dependence on both $\mum$ and $\mua$. Scale dependence is a general feature of finite-order perturbative calculations, and should be considered with care in estimating theoretical uncertainties. We have more to say on this below.

With $m_Q(\mum)$, $\alpha_s(\mua)$ related to $m_Q(m_Q)$, $\alsmz$ via RG equations, \eq{sr2} matches the general form of \eq{Olow12}, with $\alsmz$ being the only relevant element in $\Oin{k}$. There are additional inputs $\mum$, $\mua$ and $\mom_n^{Q,\text{np}}$. So in our case, \eq{Olow12} is modified as:
\bea
\mom_1^c &=& \mom_1^c \Bigl[\alsmz,m_c(m_c),\mum^c,\mua^c,\mom_1^{c,\text{np}}\Bigr],\label{M1c}\\
\mom_2^b &=& \mom_2^b \Bigl[\alsmz,m_b(m_b),\mum^b,\mua^b\Bigr]\label{M2b},
\eea
where we have neglected $\mom_2^{b,\text{np}}$. As mentioned above, the nonperturbative contribution has been claimed to be negligible for the bottom quark. We have checked this in the case of $\mom_2^b$, where the contribution from $\mom_2^{b,\text{np}}$ is below $0.1\%$, which should be compared to the experimental  uncertainty of $\mom_2^b$ of about $1\%$. Treating $\mom_1^{c,\text{np}}$ and $m_c(m_c)$ as independent inputs, which we will justify later, and focusing on the Higgs boson partial widths to $c\bar c$ and $b\bar b$ as examples of $\Ohigh$, we have, in place of \eqs{mQext}{Ohighi},
\bea
m_c(m_c) &=& m_c(m_c) \Bigl[\alsmz,\mom_1^c,\mum^c,\mua^c,\mom_1^{c,\text{np}}\Bigr],\label{mcextmu}\\
m_b(m_b) &=& m_b(m_b) \Bigl[\alsmz,\mom_2^b,\mum^b,\mua^b\Bigr],\label{mbextmu}\\
\Gamma_{H\to c\bar c} &=& \Gamma_{H\to c\bar c} \Bigl[\Oin{k},m_c(m_c),\mu_H^c\Bigr]\nonumber\\
 &=& \Gamma_{H\to c\bar c} \Bigl[\Oin{k},\mom_1^c,\mum^c,\mua^c,\mu_H^c,\mom_1^{c,\text{np}}\Bigr],\label{Gamcmu}\\
\Gamma_{H\to b\bar b} &=& \Gamma_{H\to b\bar b} \Bigl[\Oin{k},m_b(m_b),\mu_H^b\Bigr]\nonumber\\
 &=& \Gamma_{H\to b\bar b} \Bigl[\Oin{k},\mom_2^b,\mum^b,\mua^b,\mu_H^b\Bigr]\label{Gambmu},
\eea
where $\mu_H^c$, $\mu_H^b$ collectively denote other renormalization scales involved in the calculation of the partial widths. These are nevertheless not the only scale dependences for the partial widths in such an analysis. The residual scale dependences of the low-energy observables are seen to propagate into the extracted quark masses, and constitute part of the uncertainties in $m_Q(m_Q)$. These uncertainties  eventually propagate into the calculations of Higgs observables, and are reflected in the $\mum$, $\mua$ dependences in \eqs{Gamcmu}{Gambmu}. Note also that in the second equalities in \eqs{Gamcmu}{Gambmu}, the $\alsmz$ dependence in the partial widths has been changed to account for the correlation with $m_Q(m_Q)$ reflected in \eqs{mcextmu}{mbextmu}.

\eqs{Gamcmu}{Gambmu} represent the final results of the exercise of recasting Higgs observables in terms of low-energy observables, with the information contained in $m_Q(m_Q)$ fully resolved. They will be used in the next section to investigate the theoretical uncertainties in these partial widths.

To close this section we remark on the treatment of $\mom_1^{c,\text{np}}$. The known terms read~\cite{Broadhurst:1994qj}
\beq{M1cnp}
\mom_1^{c,\text{np}} = \frac{\GC}{(2m_c^{\text{pole}})^6}\biggl[-16.042-168.07\frac{\alpha_s(\mu)}{\pi} +\Ord{\alpha_s^2}\biggr],
\eeq
where $\GC$ is the gluon condensate. The commonly used value in the context of charm quark mass extraction is derived from $\tau$ decay data~\cite{Ioffe:2005ym}:
\beq{GCtau}
\GC = 0.006\pm0.012~\text{GeV}^4.
\eeq
In addition to the imprecise knowledge of $\GC$, we note two other sources of uncertainties in $\mom_1^{c,\text{np}}$. First, it is argued in~\cite{Chetyrkin:2010ic,Dehnadi:2011gc} that $\mom_1^{c,\text{np}}$ should be expressed in terms of the pole mass rather than the $\msbar$ mass in order to have a stable $\alpha_s$ expansion. We agree with this argument, but note that the use of the pole mass may introduce further ambiguities. For example, if one tries to calculate the pole mass from the $\msbar$ mass, the result will be very sensitive to the loop order. Second, considerable uncertainty is introduced by the $\mu$ dependence of the bracket in \eq{M1cnp}, since the $\Ord{\alpha_s^2}$ terms are not known. This renormalization scale is not necessarily related to $\mua$ or $\mum$ in the perturbation theory contributions [the first term in \eq{sr2}]. All these uncertainties and ambiguities will dilute any conceivable correlation between $\mom_1^{c,\text{np}}$ and $m_c(m_c)$, justifying our treatment of them as independent inputs. In our analysis the following value for $\mom_1^{c,\text{np}}$ will be assumed:
\beq{M1cnpv}
\mom_1^{c,\text{np}} = -0.0001^{+0.0006}_{-0.0014}~\text{GeV}^{-2}.
\eeq
The central value corresponds to $\GC=0.006~\text{GeV}^4$, $m_c^{\text{pole}}=1.7$~GeV and $\mu=3$~GeV in \eq{M1cnp}. The errors are very conservatively estimated by taking the extreme values $m_c^{\text{pole}}=1.4$~GeV, $\mu=1$~GeV, and varying $\GC$ in the range in~\eq{GCtau}. Even with the extreme values considered, $\mom_1^{c,\text{np}}$ is still a subpercent-level contribution to $\mom_1^c\sim0.2~\text{GeV}^{-2}$.

\section{Theoretical uncertainties of Higgs partial widths}\label{sec:higgs}

It is clear from \eqs{Gamcmu}{Gambmu} that there are two types of uncertainties in the calculation of the Higgs partial widths. Parametric uncertainty results from imprecise knowledge of the input parameters, including the input observables ($\mom_1^c$, $\mom_2^b$ and those in $\Oin{k}$) and the nonperturbative parameter $\mom_1^{c,\text{np}}$. The experimental values and errors of the input observables are:
\bea
\mom_1^c &=& 0.2121(20)(30)\text{ GeV}^{-2}~\text{\cite{Dehnadi:2011gc}},\\
\mom_2^b &=& 2.819(27)\times10^{-5}\gev^{-4}~\text{\cite{Chetyrkin:2010ic}},\\
\alsmz &=& 0.1185(6)~\text{\cite{Agashe:2014kda}},\\
m_H &=& 125.7(4)\text{ GeV}~\text{\cite{Agashe:2014kda}},\\
m_t &=& 173.21(51)(71)\text{ GeV}~\text{\cite{Agashe:2014kda}},\\
m_Z &=& 91.1876(21)\text{ GeV}~\text{\cite{Agashe:2014kda}},\\
\almz &=& 1/127.940(14)~\text{\cite{Agashe:2014kda}},\\
G_F &=& 1.1663787(6)\times10^{-5}\text{ GeV}^{-2}~\text{\cite{Agashe:2014kda}}.\label{GF}
\eea
For $\mom_1^c$ and $m_t$ the two experimental uncertainties are statistical and systematic, respectively. There is an additional systematic uncertainty in $\mom_2^b$ associated with the prescriptions used in extracting moments from data. This is discussed in~\cite{Chetyrkin:2010ic}, and we adopt ``Option A'' in that paper because among the three options considered there it appears to yield the most consistent results for $m_Q(m_Q)$ across different moments.

Perturbative uncertainty, on the other hand, is associated with unknown higher-order terms in perturbation theory calculations, and leads to residual dependence of calculated observables on the renormalization scales. When the partial widths are recast in terms of $\mom_1^c$ and $\mom_2^b$ as in \eqs{Gamcmu}{Gambmu}, multiple scales enter. $\mu_H$ comes from the calculation of the Higgs boson decay. The associated perturbative uncertainty has been studied in the literature; see e.g.~\cite{Almeida:2013jfa} where it is found to be small compared with parametric uncertainty. Here we focus on $\mum$, $\mua$, which originate from the calculation of the low-energy observables $\mom_1^c$, $\mom_2^b$ [see Eqs.~(\ref{sr2}-\ref{M2b})]. Their contribution to the total theoretical uncertainty will be singled out below by setting all input parameters to their central values in Eqs.~(\ref{M1cnpv}-\ref{GF}), and setting $\mu_H=m_H$.

\begin{figure}[t]
\flushleft\hspace{0.04in}
\includegraphics[height=3in]{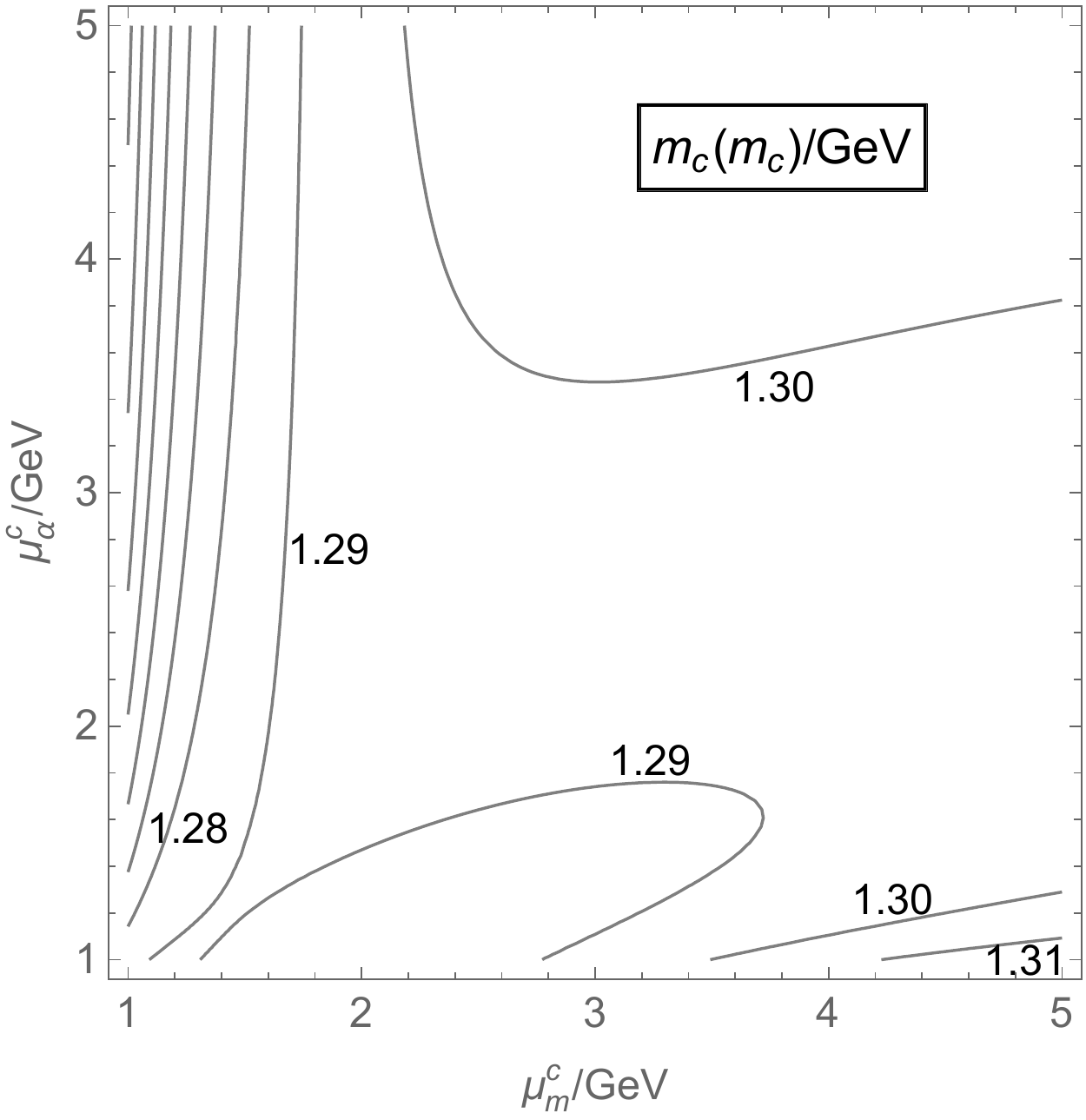}\hspace{0.3in}
\includegraphics[height=3in]{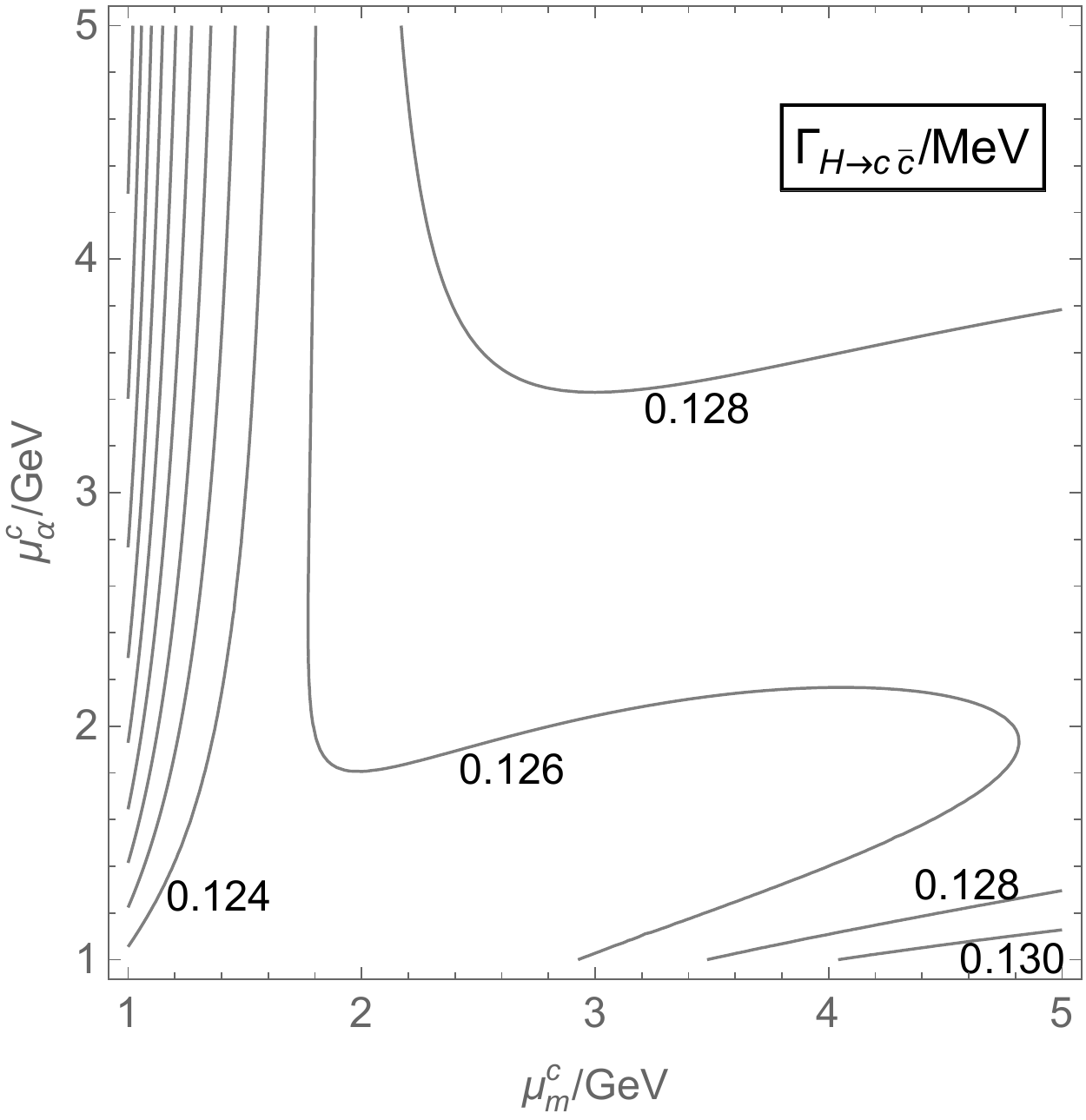}
\includegraphics[height=3.18in]{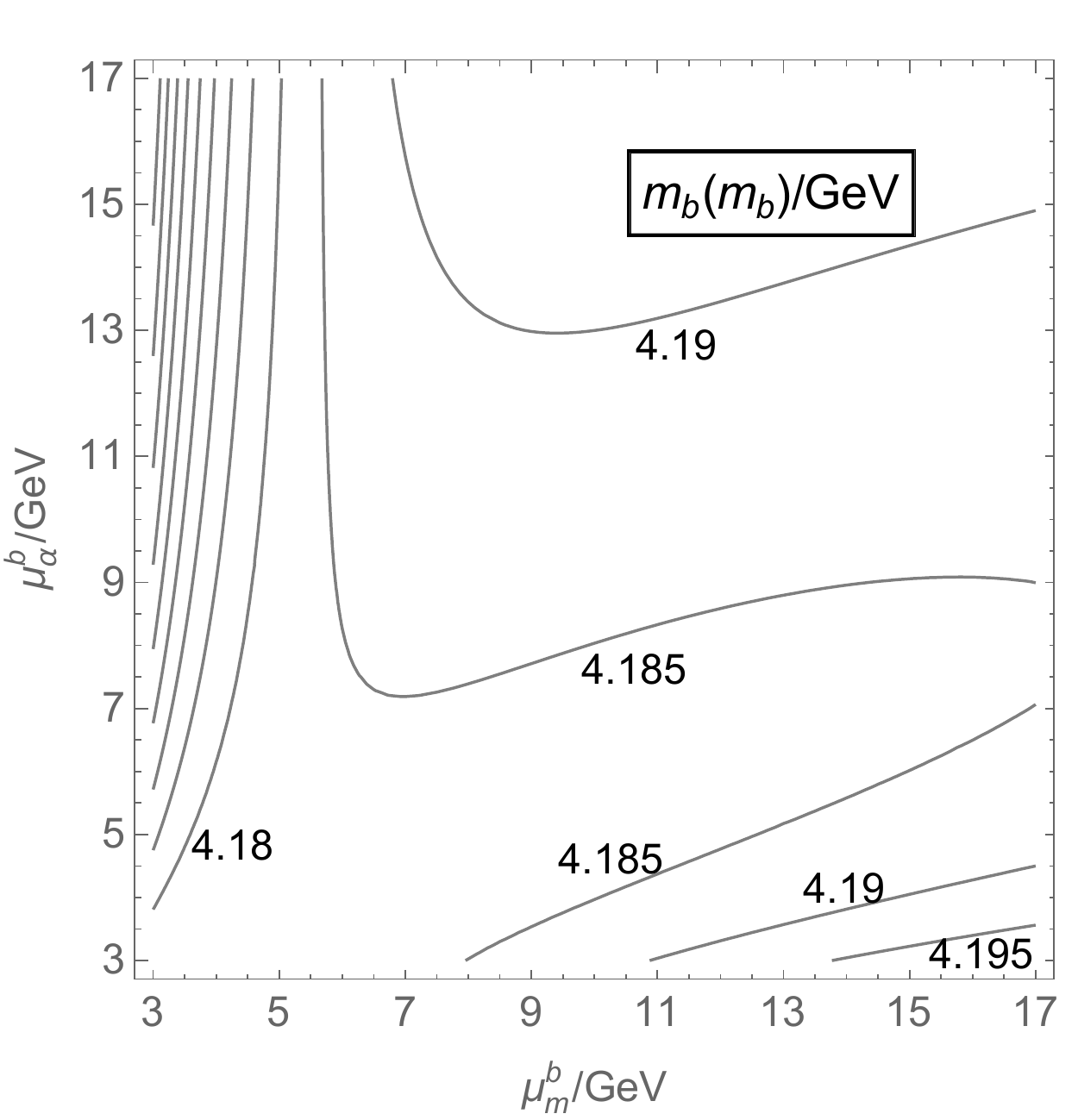}\hspace{0.15in}
\includegraphics[height=3.18in]{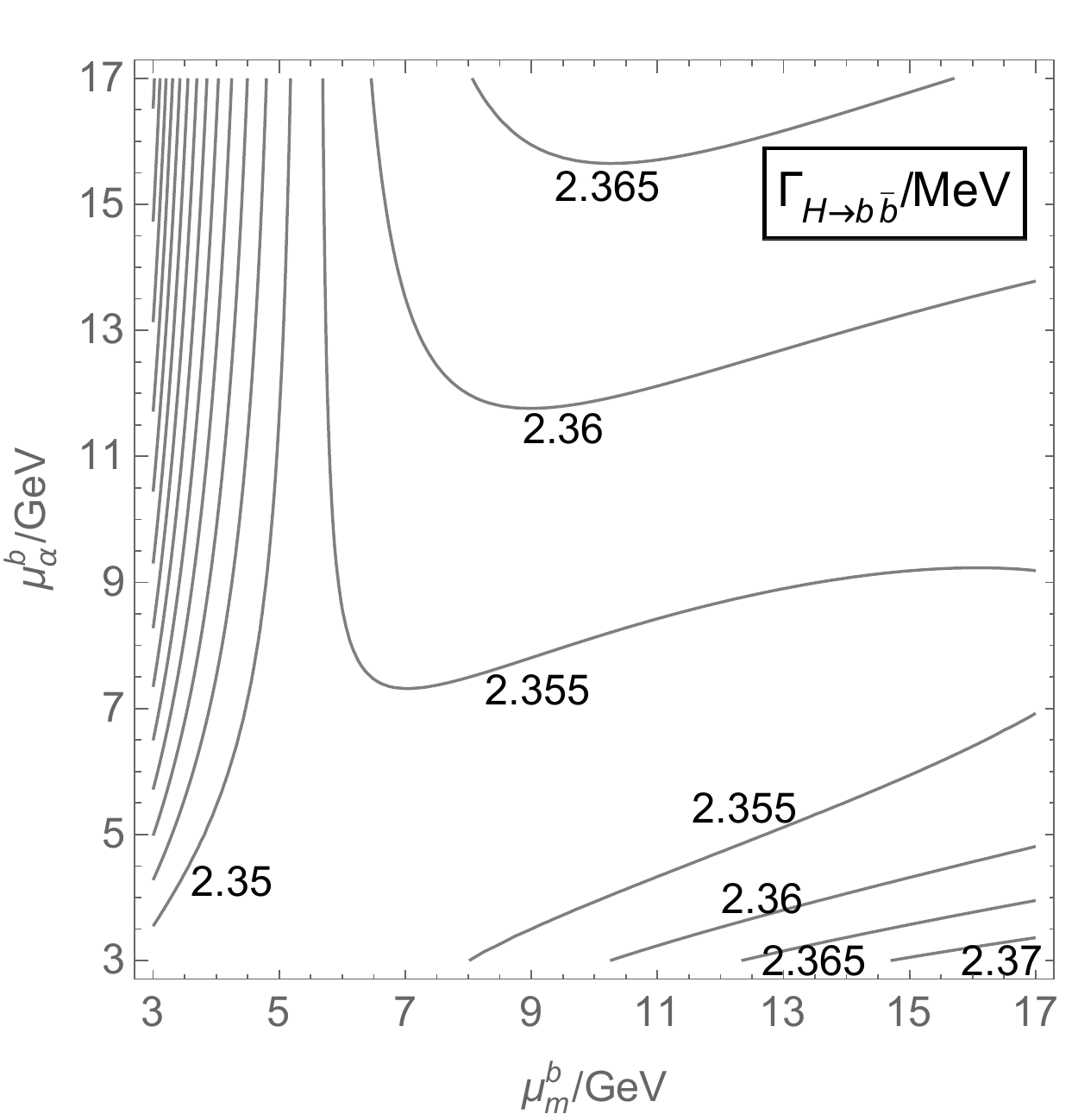}
\caption{Contours of $m_c(m_c)$ (top-left), $m_b(m_b)$ (bottom-left) in GeV, and $\Gamc$ (top-right), $\Gamb$ (bottom-right) in MeV in the $\mum$-$\mua$ plane. These plots demonstrate Eqs.~(\ref{mcextmu}-\ref{Gambmu}) with all other inputs fixed. The unlabeled contours represent decreasing values toward the top-left corner in steps of 0.01~GeV, 0.005~GeV, 0.002~MeV, 0.005~MeV, respectively.}
\label{fig:Gam}
\end{figure}

We study the perturbative uncertainty from $\mum$, $\mua$ in two steps. First, $m_Q(\mum)$ are calculated by iteratively solving \eq{sr2} following the procedure explained in~\cite{Dehnadi:2011gc}, from which $m_Q(m_Q)$ are derived. We use the \texttt{RunDec} package~\cite{Chetyrkin:2000yt} for RG running and threshold matching to the highest loop order implemented in the package. Second, the partial widths $\Gamc$, $\Gamb$ are calculated using the expansion formulas in~\cite{Almeida:2013jfa}. The results of both steps are shown in Fig.~\ref{fig:Gam} as contour plots in the $\mum$-$\mua$ plane\footnote{The numerical difference between our $m_c(m_c)$ contour plot and Fig.~6(c) in~\cite{Dehnadi:2011gc} is due to the input $\mom_1^c$ and $\alsmz$ used, and to a lesser extent the treatment of $\mom_1^{c,\text{np}}$.}. They correspond to Eqs.~(\ref{mcextmu}-\ref{Gambmu}) with other inputs fixed. These plots illustrate the propagation of $\mum$, $\mua$ dependence from low-energy moments calculations to Higgs partial widths.
%

To estimate the perturbative uncertainty, a common practice is to identify a characteristic scale of the process of interest, and vary the renormalization scale within a factor of two around that scale. For example, $\mu_H$ has been varied from $m_H/2$ to $2m_H$ in~\cite{Almeida:2013jfa}. However, this method is not directly applicable to $\mum$ and $\mua$, since $\mom_n^Q$ receive contributions from all energy scales as evident in~\eq{momdef}. One might guess from qualitative features of $R_Q(s)$ that the characteristic scale should be $\Ord{2m_Q}$, the masses of quarkonium resonances. But due to the relatively large value of $\alpha_s$ in the low-energy regime, the exact number, and hence the range in which we choose to vary $\mum$, $\mua$ can greatly affect the result of our uncertainty estimates. This is already clear from Fig.~\ref{fig:Gam}, where $\Gamc$ and $\Gamb$ are seen to exhibit rapid variation in the low-$\mum$ regime.

\begin{figure}[t]
\centering
\includegraphics[width=3.15in]{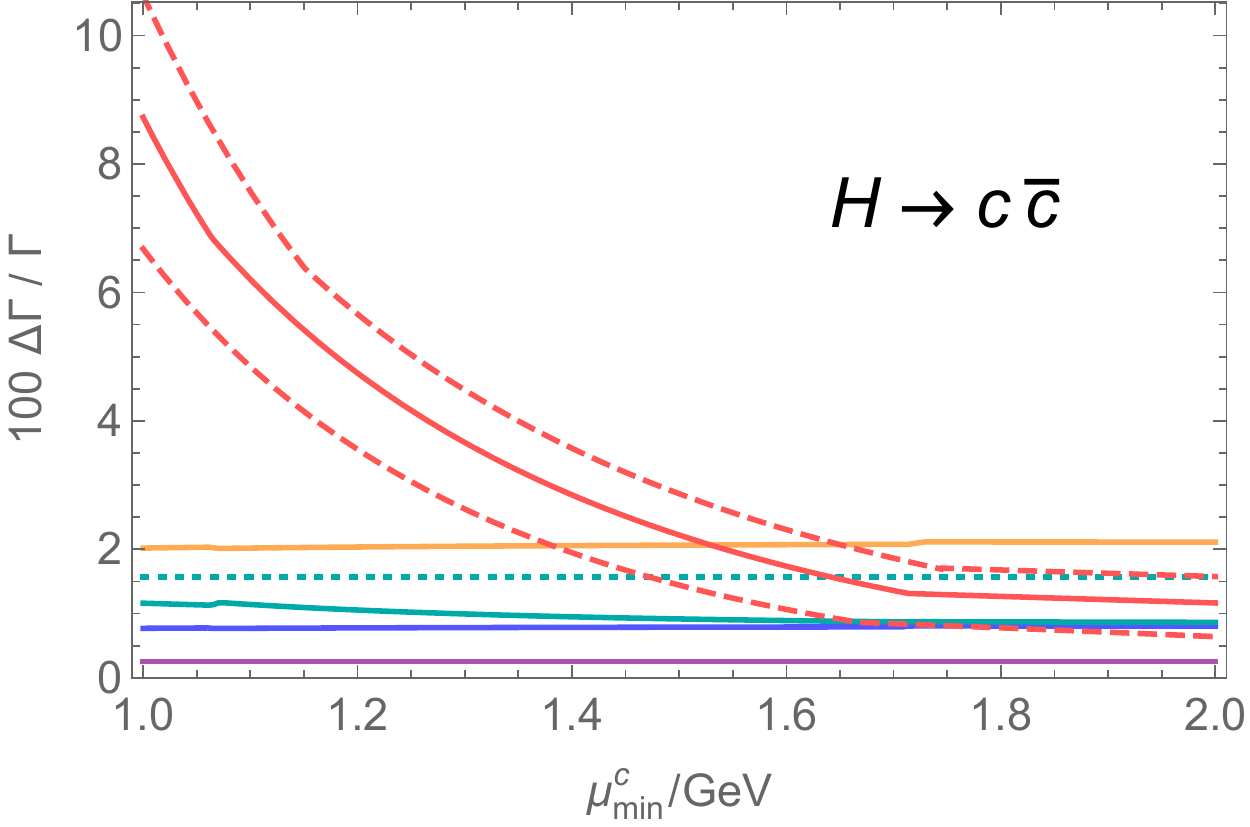}\hspace{0.1in}
\includegraphics[width=3.15in]{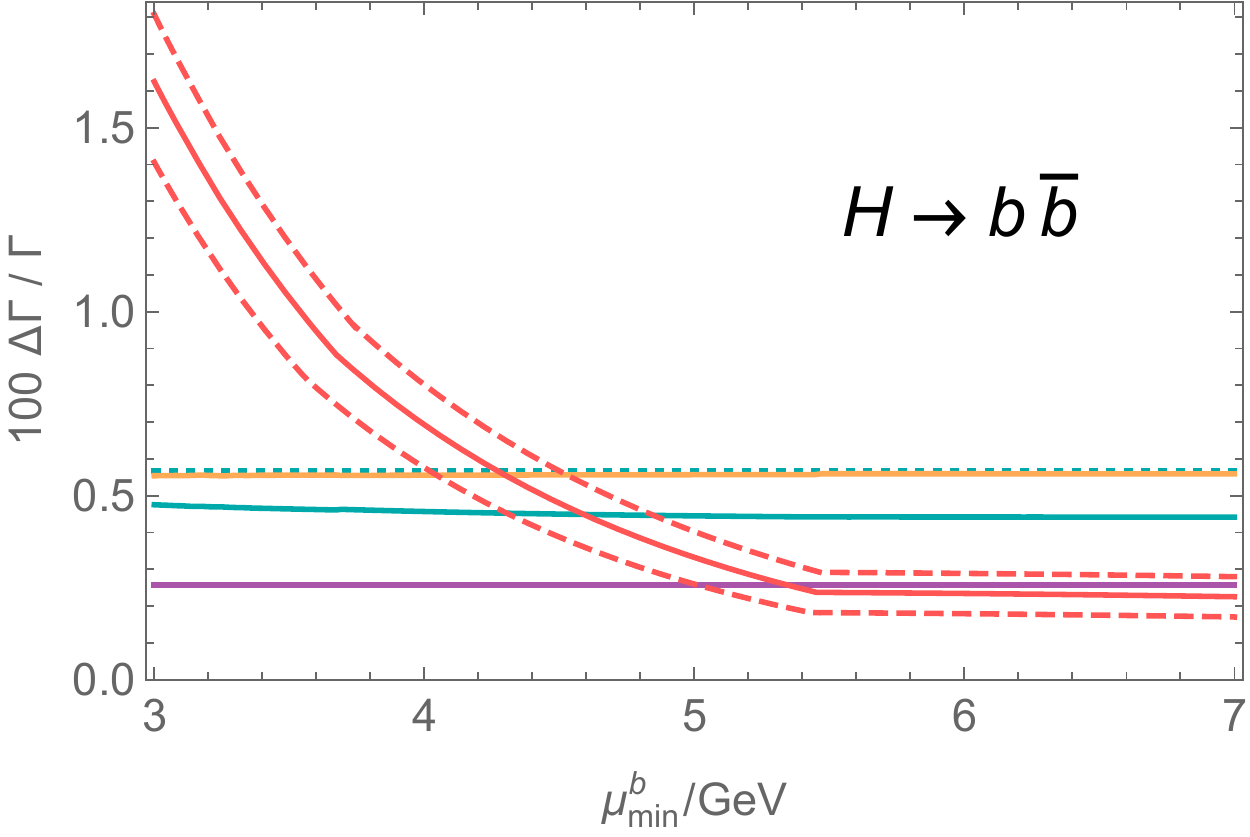}
\caption{Percent relative uncertainties in $\Gamc$ (left) and $\Gamb$ (right) as functions of $\mumin$ from various sources: perturbative uncertainty with $\mumax^c=4$~GeV, $\mumax^b=15$~GeV (red solid) or alternatively $\mumax^c=3,5$~GeV, $\mumax^b=13,17$~GeV (red dashed), parametric uncertainties from $\mom_1^c$ or $\mom_2^b$ (orange), $\alsmz$ (cyan solid), $\mom_1^{c,\text{np}}$ (blue, for $\Gamc$ only) and $m_H$ (purple). The parametric uncertainty from $\alsmz$ incorrectly calculated assuming no correlation with $m_Q$ (cyan dotted) is also shown for comparison. The parametric uncertainties are defined as shifts of the central values of $\Gamc$ and $\Gamb$ for $\mumin\le\mum,\mua\le\mumax$ caused by varying the input parameters within the errors quoted in Eqs.~(\ref{M1cnpv}-\ref{GF}), with $\mumax^c=4$~GeV, $\mumax^b=15$~GeV (the kinks are due to the maximum or minimum shifting to a different region in the $\mum$-$\mua$ plane), and are found to be insensitive to $\mumax$.}
\label{fig:PRU}
\end{figure}

Lacking an optimal method to estimate the perturbative uncertainty, we refrain from giving exact numbers, but instead aim to illustrate the ambiguity in the estimate of perturbative uncertainty by varying $\mum$ and $\mua$ independently within an adjustable range $[\mumin,\mumax]$. We will focus on the uncertainties in the partial widths, and remark that they are related to the uncertainties in $m_Q(m_Q)$ by~\cite{Almeida:2013jfa}
\be
\frac{\Delta\Gamc}{\Gamc} \simeq \frac{\Delta m_c(m_c)}{10~\text{MeV}} \times 2.1\%,\quad \frac{\Delta\Gamb}{\Gamb} \simeq \frac{\Delta m_b(m_b)}{10~\text{MeV}} \times 0.56\%.
\ee
The perturbative uncertainty, defined as half the difference between the maximum and minimum values of $\Gamc$, $\Gamb$, depends on $\mumin$ and $\mumax$. We present the results in Fig.~\ref{fig:PRU} in terms of ``percent relative uncertainties,'' defined to be $100\Delta\Gamma/\Gamma$. The red solid curves show the estimated perturbative uncertainties as functions of $\mumin$, with $\mumax^c$ ($\mumax^b$) fixed at 4 (15) GeV. Alternative choices for $\mumax^c$ ($\mumax^b$), 3 and 5 (13 and 17) GeV, give rise to the red dashed curves. These can be compared with the dominant parametric uncertainties shown by the other curves in Fig.~\ref{fig:PRU} (see figure caption for details). The popular choices in the literature $(\mumin^c,\mumax^c)=(2,4)$~GeV and $(\mumin^b,\mumax^b)=(5,15)$~GeV yield perturbative uncertainties of 1.2\% and 0.33\% for $\Gamc$ and $\Gamb$, respectively, comparable with parametric uncertainties. However, the perturbative uncertainties increase rapidly and dominate the total theoretical uncertainties if lower renormalization scales are considered. The result of the theoretical uncertainty estimate is then strongly dependent on the artificial choice of $\mumin$. This poses a serious ambiguity in precision analysis, and calls for more enlightened prescriptions for the uncertainty estimate. We note two possible directions in this regard.

The first direction was suggested very recently in~\cite{Dehnadi:2014kya} in the context of $m_Q$ extraction. There it is argued that the large perturbative uncertainty from completely uncorrelated variation of $\mum$ and $\mua$ is probably an overestimate. To get the perturbative uncertainty under control, a ``convergence test'' is performed to identify regions in the $\mum$-$\mua$ plane where the perturbative series converges too slowly (characterized by a large convergence parameter). These regions are then discarded in the uncertainty estimate. Following the approach outlined in~\cite{Dehnadi:2014kya}, we find that the discarded regions correspond to the upper-left and bottom-right corners in each plot in Fig.~\ref{fig:Gam}, where $m_Q(m_Q)$ and the partial widths exhibit rapid variation. The final result in~\cite{Dehnadi:2014kya} is a reduced perturbative uncertainty: 14 MeV and 10 MeV for $m_c(m_c)$ and $m_b(m_b)$, respectively, corresponding to 2.9\% and 0.56\% relative uncertainties in $\Gamc$ and $\Gamb$, respectively.

The convergence test is a well-motivated idea, reflecting the intuition that a proper scale choice should not lead to very slow convergence. However, further study is necessary to examine various details of the approach. For instance, one may consider loosening the constraints $m_c(m_c)\le\mum^c,\mua^c\le4~\text{GeV}$, $m_b(m_b)\le\mum^b,\mua^b\le15~\text{GeV}$ imposed in~\cite{Dehnadi:2014kya}. In particular, $\mum$, $\mua$ slightly lower than $m_Q(m_Q)$ should be allowed as long as one retains 4-flavor (5-flavor) effective strong coupling for the charm (bottom) quark. Also, the convergence criterion may be refined. The definition of the convergence parameter in~\cite{Dehnadi:2014kya} assumes an approximate geometric series behavior of the $\alpha_s$ series, but we find the latter falls off more slowly than a geometric series in most cases. Furthermore, it remains to seek a less arbitrary prescription for the fraction of $(\mum,\mua)$ to be discarded, and to investigate whether the convergence parameter is a good indicator of the size of higher-order corrections. In any case, to be conservative the reduced perturbative uncertainties mentioned above should be interpreted with caution before the approach is developed further.

As an alternative direction, one may consider the possibility of finding an optimal scale via a defensible scale-setting procedure, such as the one advocated by Brodsky-Lepage-Mackenzie (BLM)~\cite{Brodsky:1982gc}. The BLM scale for an observable is obtained by absorbing the $n_f$ terms in the perturbation series, which come from the QCD beta function, into the running coupling $\alpha_s$. This is arguably the physical scale of the process, with higher-order corrections associated with RG running appropriately resummed. We also note that the BLM procedure extended to all orders based on the principle of maximum conformality~\cite{Brodsky:2013vpa} has been demonstrated to be self-consistent~\cite{Wu:2013ei}. In the case of $\mom_n^Q$, however, there are two renormalized parameters $\alpha_s$ and $m_Q$, and naive application of the BLM procedure might be problematic. This is because even when the $n_f$ terms are absorbed into running $\alpha_s$ and/or $m_Q$, the leading-order mass renormalization, which is independent of $n_f$, may lead to large loop corrections which are difficult to identify. Indeed, we find that naive application of BLM, namely absorbing the $n_f\alpha_s^2$ terms, sets scales for $\mum$ and $\mua$ which are strongly disfavored by the convergence test. In light of the importance of a more precise $m_Q$ determination, it might be worthwhile to investigate the nontrivial possibility of generalizing the BLM method and its extensions~\cite{Brodsky:2013vpa,Kataev:2014jba} to include running quark masses.

The parametric uncertainties, on the other hand, are seen from Fig.~\ref{fig:PRU} to be dominated by experimental measurement uncertainties of $\mom_1^c$ and $\mom_2^b$ (orange). Reduction of these will rely on more precise measurements of $R_Q(s)$ and more careful treatment of experimental data. At present the major problem is the lack of data above $\sqrt{s}=11.2$~GeV, resulting in large uncertainties in the bottom quark moments~\cite{Chetyrkin:2010ic}. Also, the quarkonium resonances are currently treated in the narrow width approximation, the quality of which should be examined in light of higher precision requirements in the future. $\alsmz$ (cyan solid) constitutes a subdominant source of parametric uncertainties. Its contribution is seen to be smaller than the incorrect estimate assuming no correlation between $\alsmz$ and $m_Q$ (cyan dashed), due to partial cancelation between direct $\alsmz$ dependence and indirect dependence through $m_Q$. With our conservative  estimate (i.e.~erring on the large side) in \eq{M1cnpv}, $\mom_1^{c,\text{np}}$ leads to an uncertainty in $\Gamc$ (blue) at a similar level as $\alsmz$. This may represent a challenge in the future, and calls for further investigation of the gluon condensate contribution. The uncertainty due to $m_H$ (purple) is less important, while other input observables listed at the beginning of this section have a negligible effect on the parametric uncertainty.

\section{Conclusions}

For the precision Higgs program to succeed in the future, additional effort is required to improve the precision of SM calculations in order to match the proposed experimental accuracy. A better understanding of theoretical uncertainties is critical. Toward this aim, we emphasize the role of low-energy observables, and further propose the idea of a global analysis incorporating relevant observables across all energy regimes. Rather than contributing indirectly via the charm and bottom quark masses, low-energy observables explicitly participate in such a precision analysis. Future studies in this direction should examine all candidates of low-energy observables, and determine an efficient set of observables for the global fit.

In the context of precision Higgs calculations, we focused on the Higgs boson partial widths to charm and bottom quarks, and investigated the theoretical uncertainties in these observables. By eliminating charm and bottom quark masses in favor of low-energy observables $\mom_1^c$ and $\mom_2^b$, we recast the partial widths in terms of these and other input observables. Much information originally hidden in uncertainties in the highly processed quark masses becomes transparent. Experimental uncertainties in the low-energy observables are directly propagated into the Higgs partial widths, and the uncertainty due to $\alsmz$ is treated properly. Perturbative uncertainties are difficult to assess due to the ambiguity in the choice of renormalization scales in the low-energy regime, and can dominate the total theoretical uncertainty of the Higgs partial widths if lower values of the renormalization scales are considered than is usually the case in the literature.

Such analysis points to future directions in the precision program. For the partial widths considered here, we note that while future experimental progress could potentially reduce parametric uncertainties significantly, our ability to make precise predictions on the Higgs partial widths will not improve unless better understanding of the perturbative uncertainty is achieved. As for $\mom_1^c$ and $\mom_2^b$ studied here, this might require the calculation of $\Ord{\alpha_s^4}$ corrections to $\Pi_Q(q^2)$ (in the low-$q^2$ limit) and/or more enlightened scale setting. Though the actual situation may be better in a global fit where $\mom_1^c$ and $\mom_2^b$ are not the only low-energy observables involved, it remains crucial to carefully investigate whether the scale-setting problem is also present for other low-energy observables sensitive to $m_Q$. If the perturbative uncertainty gets under control, the precision program, where both low-energy observables and Higgs observables play an important role, will be promising in studying properties of the Higgs boson, and even more generally testing the SM across a wide range of energy scales and probing new physics ideas.

\acknowledgments

We thank J. Shigemitsu for useful discussions.
J.D.W.\ and Z.Z.\ are supported in part by DOE under grant DE-SC0011719.
S.P.\ is supported by the National Science Center in Poland under the research grants  DEC-2012/05/B/ST2/02597 and DEC-2012/04/A/ST2/00099.
A.A.P.\ is grateful to Fermilab's Theory Group for their hospitality.
A.A.P.\ is supported in part by the U.S. Department of Energy under contract DE-SC0007983,
Fermilab's Intensity Frontier Fellowship and URA Visiting Scholar Award \#14-S-23.
Fermilab is operated by Fermi Research Alliance, LLC, under contract DE-AC02-07CH11359 with the United States 
Department of Energy.


\end{document}